\documentclass[a4paper]{article}

\usepackage{INTERSPEECH2022,multirow}

\title{Online Target Speaker Voice Activity Detection for Speaker Diarization}
\name{Weiqing Wang$^1$, Qingjian Lin$^3$, Ming Li$^{1,2,}$\sthanks{\ \ Corresponding author: Ming Li.}}
\address{
  $^1$Department of Electrical \& Computer Engineering, Duke University, Durham, NC 27708, USA\\
  $^2$Data Science Research Center, Duke Kunshan University, Kunshan 215316, PR China\\
  $^3$AI Lab, Lenovo Research, Beijing 100085, PR China }
\email{\{weiqing.wang, ming.li369\}@duke.edu}

\begin{document}
%\ninept

\maketitle
\begin{abstract}
This paper proposes an online target speaker voice activity detection system for speaker diarization tasks, which does not require a priori knowledge from the clustering-based diarization system to obtain the target speaker embeddings. First, we employ a ResNet-based front-end model to extract the frame-level speaker embeddings for each coming block of a signal. Next, we predict the detection state of each speaker based on these frame-level speaker embeddings and the previously estimated target speaker embedding. Then, the target speaker embeddings are updated by aggregating these frame-level speaker embeddings according to the predictions in the current block. We iteratively extract the results for each block and update the target speaker embedding until reaching the end of the signal. Experimental results show that the proposed method is better than the offline clustering-based diarization system on the AliMeeting dataset. 
\end{abstract}

\section{Introduction}
Speaker diarization is the task that assigns speaker labels to speech regions, which is a very important pre-processing pipeline for many downstream tasks and improves their performance, e.g., multi-speaker automatic speech recognition~\cite{ASR1, ASR2}.

For the conventional clustering-based diarization system, it is still challenging to recognize the overlapping regions under the scenarios of meeting, interview, and telephone. The reason is that these scenarios contain lots of overlapping speech, but the conventional clustering-based system cannot directly detect these overlapping regions without the help of additional modules. Recently, some neural post-processing techniques have been proposed to reduce the miss error brought by the overlapping speech, including overlapping speech detection~\cite{BUTdihard2}, target speaker voice activity detection (TS-VAD)~\cite{TSVAD}, and end-to-end neural diarization (EEND) as post-processing~\cite{EENDPost}. The EEND was first proposed to directly obtain the overlap-aware results from the speech signal, which has shown superior performance on some datasets and challenges than the clustering-based results~\cite{EEND1, EEND2, EEND3, EEND4}. 

Although the aforementioned neural strategies have shown significant improvement over the conventional diarization system on multiple benchmark databases, most of them only work in an offline manner. Recently, several online methods have been proposed for speaker diarization ~\cite{patino18_odyssey, geiger2010gmm, markov2008improved, UISRNN2019, fini2020supervised, zhu2016online, dimitriadis2017developing}. In addition, many end-to-end methods have been extended in an online manner. Inspired by the EEND with encoder-decoder-based attractors (EDA)~\cite{EEND3}, Eunjung et al.~\cite{BW-EDA-EEND} introduced a family of block-wise (BW) EDA-EEND (BW-EDA-EEND) that enables online speaker diarization. Yawen et al.~\cite{OnlineEEND} also proposed another online EDA-EEND model with speaker tracking buffer (STB) that reduces the chunk size to 1 second.

In this paper, we propose an online TS-VAD method to extend the original TS-VAD into an online manner. TS-VAD has shown an excellent performance in many challenges in recent years~\cite{TSVAD, wang21y_interspeech, wang2021dku}. For the original TS-VAD, we first obtain the target speaker embedding using a clustering-based diarization system and then feed these embeddings to the TS-VAD model to predict the refined diarization results. The reason that TS-VAD has superior performance is that it can retrieve the speaker identity from an overlapping region given the corresponding target speaker embedding. 

Since the original TS-VAD is performed in a block-wise manner, it can also be considered an online diarization system if given the target speaker embeddings. However, the target speaker embeddings are always obtained from the offline diarization system. If we also obtain these embeddings in an online manner, this TS-VAD model can become an online diarization system. To cope with this, we employ a ResNet-based front-end model to extract the frame-level embeddings that are not only for target speaker detection but also for target speaker embedding aggregation. This means that the TS-VAD model and the speaker embedding extractor share the same front-end model. The target speaker embeddings start from several zero vectors, and they are updated given the frame-level embeddings and estimated results of each coming block. Since the proposed online TS-VAD model generates the diarization results from scratch and it does not require an extra speaker embedding extractor, it can also be considered as an end-to-end model. In addition, the computational cost is significantly reduced since we do not need to extract the speaker embedding with another model. 

%The rest of this paper is organized as follows: Section \ref{sec:ts-vad} introduces our online TS-VAD system. Experimental details and results are presented in Section \ref{sec:exp}. Section \ref{sec:conclusion} concludes this paper. 

\begin{figure*}[h]
  \centering
  \includegraphics[scale=1.1]{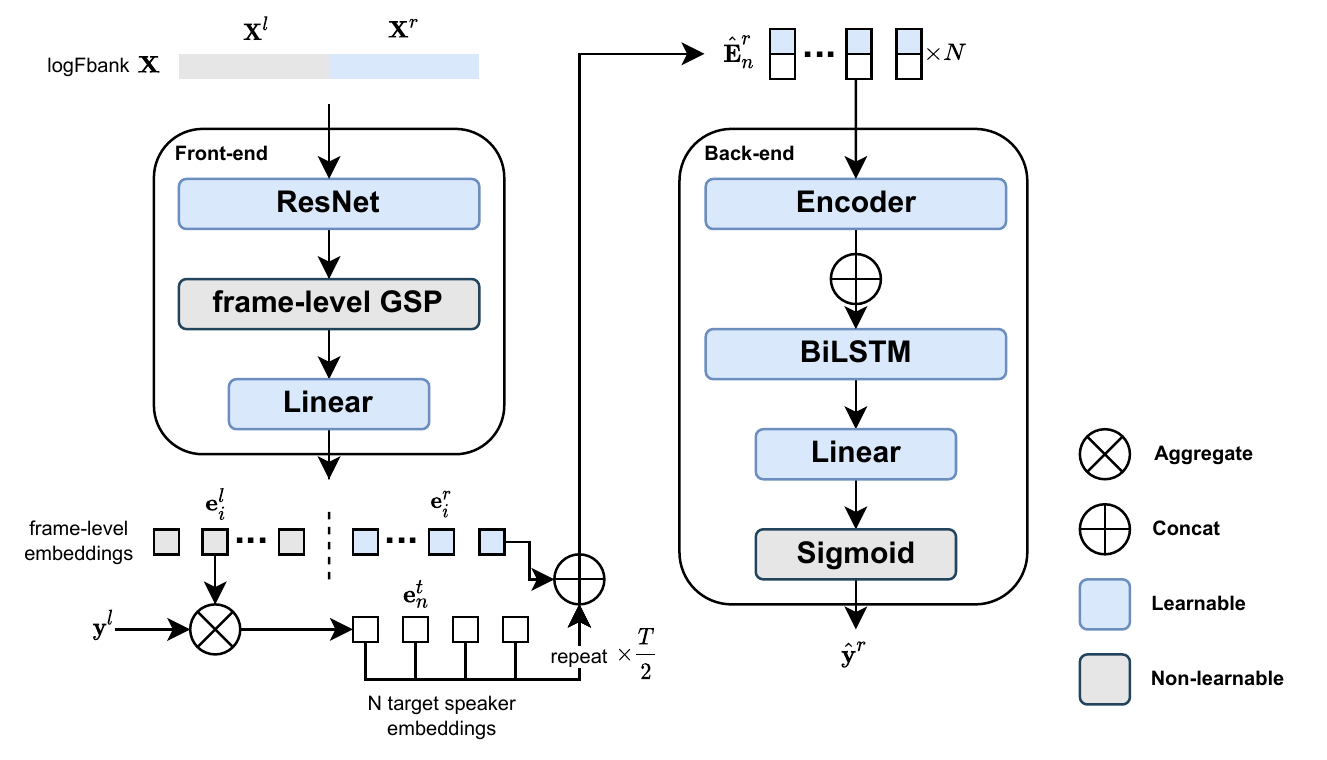}
  \caption{The online TS-VAD model architecture and training process for single-channel data.}
  \label{fig:train}
\end{figure*}

\section{Proposed Online TS-VAD}
\label{sec:ts-vad}

\subsection{Model Architecture}
The model contains a ResNet-based front-end module and a self-attention based back-end module, as shown in Figure \ref{fig:train}. For the front-end module, we employ a ResNet34 with a linear layer that converts the acoustic features to the frame-level embeddings. For the back-end module, we employ an encoder to separately extract the decision state from the concatenated speaker embeddings. Later, given the concatenation of the decision state from $N$ different target speakers, a BiLSTM and a linear layer are adopted to predict the final diarization results. The model architecture is the same as that of our offline TS-VAD model in \cite{wang2022cross}, and we only change the training and inference process to obtain the target speaker embedding in an online manner. We will discuss the details of training and inference process in Section \ref{sec:training} and \ref{sec:inference}, respectively.

\subsection{Training process}
\label{sec:training}

For the front-end, given an acoustic feature $\mathbf{X}\in\mathbb{R}^{F \times L}$, the ResNet extract a temporal feature map $\mathbf{M} \in \mathbb{R}^{C \times \frac{F}{8} \times \frac{L}{8}}$, where $C=256$ is the number of channels, F is the dimension of features, and L is the number of frames.  For the global statistic pooling (GSP) layer, we take the mean and standard deviation of the feature map in each channel, producing a $2C$ dimensional vector. To extract this vector in frame-level, we perform the GSP for each frame of the feature map $\mathbf{M}$, producing 
$T$ vectors with a size of $2C$, where $T = \frac{L}{8}$. Finally, the linear layer outputs the speaker embedding for each frame.

After we obtain the frame-level speaker embedding $\mathbf{E}=\left[\mathbf{e}_1, ..., \mathbf{e}_T\right] \in \mathbb{R}^{T \times D}$, we equally split $\mathbf{E}$ into two sub-sequences, where the left is $\mathbf{E}^l=\left[\mathbf{e}^l_1, ..., \mathbf{e}^l_i, ..., \mathbf{e}^l_{\frac{T}{2}}\right]$ and the right is $\mathbf{E}^r=\left[\mathbf{e}^r_1, ..., \mathbf{e}^r_j, ..., \mathbf{e}^r_{\frac{T}{2}}\right]$. Next, we aggregate the left part $\mathbf{E}^l$ to obtain the target speaker embedding using the 
corresponding label $\mathbf{y}^l\in\{0, 1\}^{\frac{T}{2} \times N}$ by:

\begin{align}
    \label{eq:aggregate}
    \mathbf{e}^t_n = \frac{\mathbf{y}^l_n \odot \mathbf{E}^l}{\sum_{i=1}^\frac{T}{2}(\mathbf{y}^l_{i, n})}, n \in \{1, 2, ..., N\}
\end{align}
where i is the index of frames, n is the index of speakers, $N$ is the number of target speakers, and $\mathbf{e}^t_n$ is the target speaker embedding of the $\text{n}^\text{th}$ speakers. Actually, equation (\ref{eq:aggregate}) takes the mean of all frames of the $\text{n}^\text{th}$ speaker's frame-level embedding. If the $\text{n}^\text{th}$ speaker does not appear in the label $\mathbf{y}^l$, which means that the denominator is 0, we directly set $\mathbf{e}^t_n$ as a zero vector. 

Finally, the $\text{n}^\text{th}$ target speaker embedding $\mathbf{e}^t_n$ will be repeated $\frac{T}{2}$ times and concatenated with the right frame-level embeddings $\mathbf{E}^r$, producing a concatenated embedding $\hat{\mathbf{E}}^r_n \in \mathbb{R}^{\frac{T}{2} \times 2D}$. This is the input of the back-end module. 

For the back-end module, it is the same as that of the original TS-VAD model. The Encoder separately processes $\hat{\mathbf{E}}^r_n$ for each $\text{n}^\text{th}$ speaker, and then produce the frame-level detection states for each speaker. Later, these detection states are concatenated again, and a BiLSTM layer with linear layer produces the presence probabilities $\hat{\mathbf{y}^r}\in\mathbb{R}^{\frac{T}{2} \times N}$ of all speakers. Finally, we employ binary cross-entropy (BCE) to calculate the loss between the outputs and labels. 

\subsection{Inference process}
\label{sec:inference}
\begin{figure}[h]
  \centering
  \includegraphics[width=\linewidth]{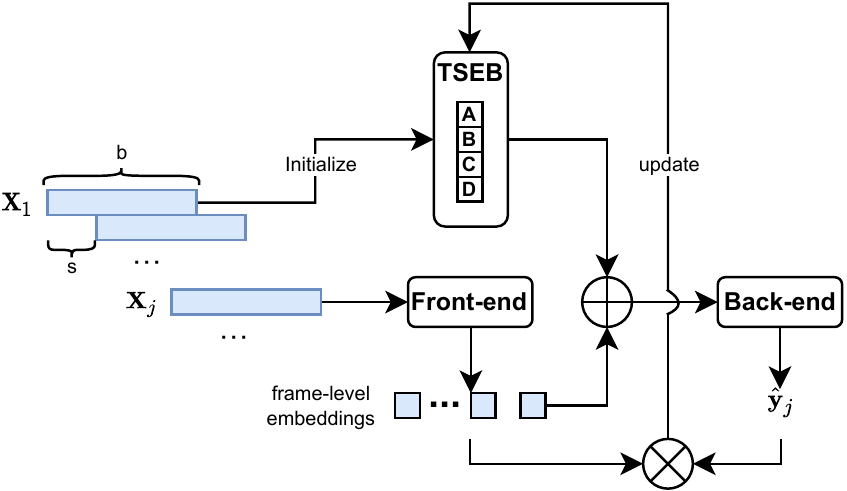}
  \caption{The inference process.}
  \label{fig:train}
\end{figure}

In the training process, the silence regions are removed in advance to ensure that the model focuses on learning the difference between speakers. Therefore, in the inference stage, we also need an additional VAD model to identify the speech frames. Once we obtain enough speech frames that form a block, the TS-VAD model can take this block as input and produce the diarization results for this block. For convenience, we use the oracle VAD to remove the silent frames. 

The inference is performed in a block-wise manner, where the block size is $b$ seconds, the block shift is $s$ seconds and $b=ks$. We create a target speaker embedding buffer (TSEB) that stores the mean of the frame-level embeddings and the corresponding number of frames for each speaker, which is initialized by the first block. For initialization, we assume that the signal of the first $s$ seconds only contains one speaker, and we obtain the target speaker embedding from the first block in the following steps:
\begin{enumerate}
    \item Set $k^{'}=1$. Directly take the mean of the frame-level embeddings of the first $s$ seconds as the target speaker embedding of the first speaker, and other target speaker embeddings are set to zero vectors. Also record the number of frames of the first embedding.
    \item $k^{'}=k^{'}+1$. Start to infer for the signal of the first $k^{'}s$ seconds and obtain the output $\hat{\mathbf{y}}\in\mathbb{R}^{T \times N}$. If there exist some frames that the existence probabilities of all N speakers are lower than a preset threshold $t_\text{init}$, we believe that these frames belong to a new speaker, and the corresponding frames in $\hat{\mathbf{y}}$ will be set to values of 1. Then, we update the TSEB using the binary decision of the output $\hat{\mathbf{y}}$ and the frame-level embeddings of the first $k^{'}s$ seconds with equation \ref{eq:aggregate}. In addition, the number of frames of each speaker is also updated.
    \item If $k^{'} <= k$, go back to step 2. Otherwise, stop iteration and the target speaker embeddings obtained from the first block (the first $ks$ seconds) will be used for inference of the subsequent blocks.  
\end{enumerate}

Next, the initialized target speaker embeddings in TSEB are the input for the next block. Since we set the block shift $s$ small enough, we can assume that there should be at most one new speaker in the next $s$ seconds. After extracting the frame-level embeddings from this block, we can obtain the output $\hat{\mathbf{y}}$ with the initialized TSEB. To update the target speaker embeddings, three different thresholds are employed:
\begin{itemize}
    \item Lower threshold $t_\text{low}$: If all N existence probabilities of a frame are lower than this threshold, assign a new speaker for this frame. 
    \item Upper threshold $t_\text{up}$: Only update the TSEB with the frames whose existence probabilities are greater than this threshold. (for higher purity)
    \item Decision threshold $t_\text{d}$: The threshold that used to obtain the final diarization results from output $\hat{\mathbf{y}}$. Usually set to 0.5. 
\end{itemize}

First, we compare $\hat{\mathbf{y}}$ with $t_\text{low}$. If a new speaker is found, we assign the frames of the new speaker in $\hat{\mathbf{y}}$ with values of 1. Next, we compare the $\hat{\mathbf{y}}$ with $t_\text{up}$ and obtain the binary decision, where the overlapping frames are excluded. Then we update TSEB with this binary decision and the frame-level embeddings of this block. Since we have recorded the mean of the embeddings and the number of frames, the TSEB can be updated in the following steps:

\begin{enumerate}
    \item For each speaker, get the sum of all frame-level embeddings from TSEB given the mean and the number of all frames. 
    \item For each speaker, the sum of frame-level embeddings can be updated by adding the corresponding frame-level embeddings in the current block according to the binary decision of $\hat{\mathbf{y}}$. 
    \item Finally, update the mean of each target speaker embedding and the corresponding number of frames according to the binary decision of $\hat{\mathbf{y}}$.
\end{enumerate}

Each time a new speaker is found, we will extract the output again for the current block with the updated TSEB. If TSEB has already contains $N$ speakers, we will not add a new speaker. The inference and TSEB updating process can be iteratively performed until reaching the end of the signal.

\begin{table*}[htbp] 
  \caption{The DERs (\%) of different systems on AliMeeting Eval and Test set. Both the official baseline and winner system are offline. The online TS-VAD is performed on the first channel of the signal with $b=16$ and $s=2$.}
  \label{tab:results}
  \centering
  \begin{tabular}[c]{lccccccccc}
    \toprule
     \multirow{2}*{\textbf{Model}} &  \multicolumn{4}{c}{\textbf{AliMeeting Eval set}} &  \multicolumn{4}{c}{\textbf{AliMeeting Test set}}  \\
     
     \cmidrule(lr){2-5} \cmidrule(lr){6-9}
   
      & \textbf{2 spk} & \textbf{3 spk} & \textbf{4 spk} & \textbf{Total}  & \textbf{2 spk} & \textbf{3 spk} & \textbf{4 spk} & \textbf{Total} \\
   \midrule
    Official offline baseline \cite{Yu2022M2MeT}& 4.84 & 10.90 & 23.18 & 15.24 & 3.00 & 6.36 & 30.59 & 15.60 \\
    Winner system \cite{wang2022cross} \\
    \ \ \ \ Offline single-channel clustering & 5.00 & 10.06 & 20.54 & 13.79 & 3.25 & 6.79 & 27.85 & 14.54\\
    \ \ \ \ Offline single-channel TS-VAD & 0.89 & 6.63 & 5.47 & 4.11 & - & - & - & - \\
    \ \ \ \ Offline multi-channel TS-VAD & 0.61 & 3.16 & 3.05 & \textbf{2.25} & 0.15 & 5.19 & 4.37 & \textbf{2.98} \\
    \midrule
     Online single-channel TS-VAD  & 1.90 & 8.36 & 12.12 & \textbf{8.14} & 5.17 & 12.01 & 16.53 & \textbf{11.42} \\ 
     \bottomrule
     \end{tabular}

\end{table*}

% \begin{table*}[htbp] 
%   \caption{The DERs (\%) of different systems on AliMeeting Eval and Test set. Both the official baseline and winner system are offline. The online TS-VAD is performed on the first channel of the signal with $b=16$ and $s=2$.}
%   \label{tab:results}
%   \centering
%   \begin{tabular}[c]{lccccccccc}
%     \toprule
%      \multirow{2}*{\textbf{Model}} &  \multicolumn{4}{c}{\textbf{AliMeeting Eval set}} &  \multicolumn{4}{c}{\textbf{AliMeeting Test set}}  \\
     
%      \cmidrule(lr){2-5} \cmidrule(lr){6-9}
   
%       & \textbf{2 spk} & \textbf{3 spk} & \textbf{4 spk} & \textbf{Total}  & \textbf{2 spk} & \textbf{3 spk} & \textbf{4 spk} & \textbf{Total} \\
%   \midrule
%     Official modular SD & 4.84 & 10.90 & 23.18 & 15.24 & 3.00 & 6.36 & 30.59 & 15.60 \\
%     Winner system  \\
%     \ \ \ \ Offline single-channel target-segment detection & 5.00 & 10.06 & 20.54 & 13.79 & 3.25 & 6.79 & 27.85 & 14.54\\
%     \ \ \ \ Offline single-channel target-speaker detection & 0.89 & 6.63 & 5.47 & 4.11 & - & - & - & - \\
%     \ \ \ \ Offline multi-channel target-speaker detection & 0.61 & 3.16 & 3.05 & \textbf{2.25} & 0.15 & 5.19 & 4.37 & \textbf{2.98} \\
%     \midrule
%      Online single-channel TS detection  & 1.90 & 8.36 & 12.12 & \textbf{8.14} & 5.17 & 12.01 & 16.53 & \textbf{11.42} \\ 
%      Online multi-channel TS detection  & 0.84 & 4.74 & 9.62 & \textbf{5.96} & 0.33 & 15.36 & 5.50 & \textbf{5.54} \\ 
%      \bottomrule
%      \end{tabular}
% \end{table*}

\section{Experimental Results and Discussion}
\label{sec:exp}
\subsection{Dataset and Simulation}
The experiments are conducted on the first channel of the AliMeeting dataset \cite{Yu2022M2MeT}. This multi-channel dataset contains 118.75 hours of speech data, and it is divided into 104.75 hours for training, 4 hours for evaluation, and 10 hours for testing. The number of participants within one meeting session ranges from 2 to 4. The duration of each session is about 30 minutes. The average speech overlap ratio is 42.27\% for the training set and 34.76\% for the evaluation set, respectively. More details can be found in \cite{Yu2022M2MeT}.

We create a simulated dataset from the AliMeeting Train set, and the simulation process is as follows:

\begin{enumerate}
    \item As each speaker in the AliMeeting dataset has a unique identification, we select all non-overlapped speech for each speaker from the AliMeeting Train set for simulation. 
    \item Extract the labels from the transcript of the AliMeeting Train set and remove all silence regions.
    \item During the training stage, the simulated data is generated on-the-fly in an online manner, where we randomly choose a segment of the label and fill the active region with the continuous non-overlapped speech segments.
\end{enumerate}

Finally, the AliMeeting Eval set is adopted for validation, and the Test set is employed for testing. 

\subsection{Experimental Details}
The architecture of the front-end module is the same as that in \cite{cai2020within} except that we double the channel size. For the back-end, the Encoder is a 2-layer 4-head Transformer Encoder with a 1024 dimensional feed-forward layer, and the dropout is set to 0.1. The hidden size of the single BiLSTM layer is 128. The fully-connected layer projects the output from BiLSTM to a $N$-dimensional vector, where $N$ is the number of target speakers. In our experiment, $N$ is set to 4 as the maximum number of speakers is 4. 

The model is directly trained on the simulated dataset for 20 epochs, where each epoch includes 25,600 simulated samples with a batch size of 16 and a learning rate of 0.0001. Later, we continue to train the model on the AliMeeting training set for 20 more epochs with a learning rate of 0.00001. The model is optimized by Adam with BCE loss. The input is a continuous 32-second speech signal, which means that the length of both $\mathbf{X}^l$ and $\mathbf{X}^r$ is 16 seconds. The acoustic feature is the 80-dimensional log Mel-filterbank energy with a frame length of 25ms and a frame shift of 10ms. 

Due to the limit of the GPU memory, we cannot train the model with a very long signal. However, this will lead to a problem that $\mathbf{X}^l$ only contains one or two speakers with a short duration, and the back-end module only accepts one or two non-zero target speaker embeddings during the training stage. Therefore, our model only shows good performance for the 2-speaker sessions at first. To cope with this, we replace the $\mathbf{X}^l$ with a randomly simulated signal that contains three or four speakers, and the probability of replacing is set to 0.5. 

During inference stage, we test different block size $b$ and block shift $s$. For the threshold, $t_\text{init}=0.5$, $t_\text{low}=0.4$, $t_\text{up}=0.7$ and $t_\text{d}=0.5$. We only tune the $t_\text{low}$ and $t_\text{up}$ on the Eval set with grid search, whereas $t_\text{init}$ and $t_\text{d}$ are set to 0.5 intuitively.

We use the Diarization Error Rate (DER) as the evaluation metric, where a forgiveness collar of 0.25 is employed.

\begin{figure}[h]
  \centering
  \includegraphics[width=\linewidth]{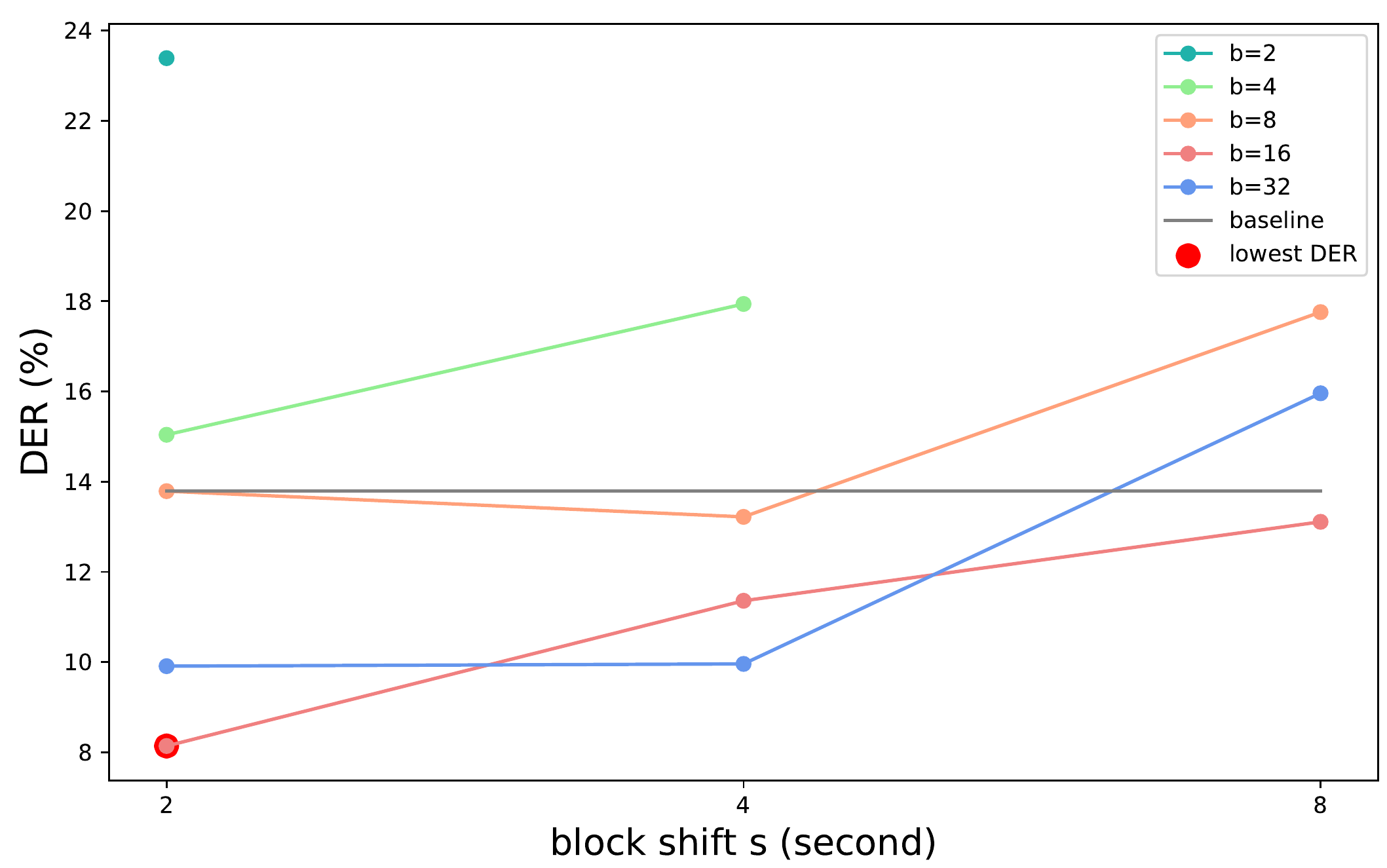}
  \caption{The DERs (\%) of the online TS-VAD system with different blocks size $b$ and block shift $s$ on AliMeeting Eval set.}
  \label{fig:results}
\end{figure}

\begin{figure}[h]
  \centering
  \includegraphics[width=\linewidth]{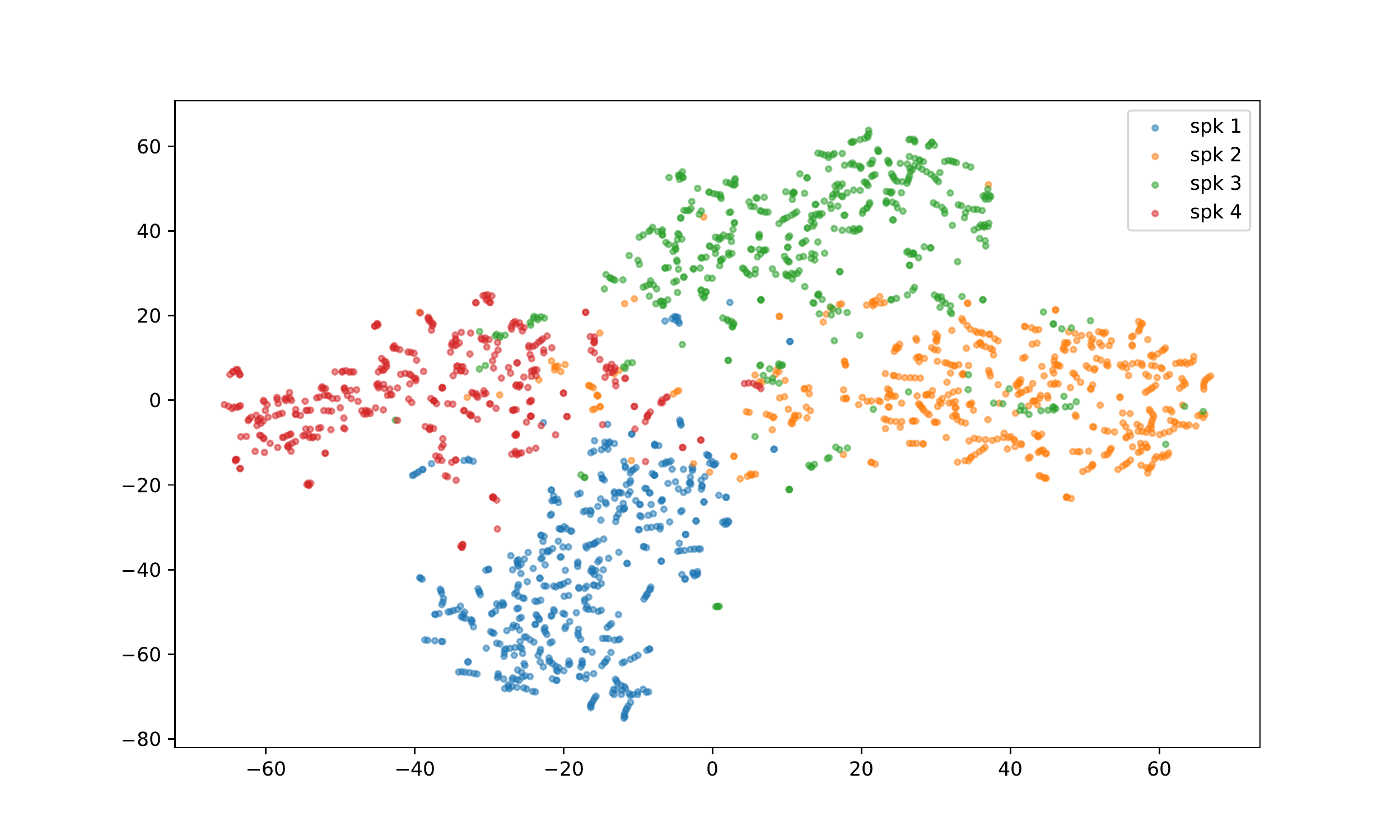}
  \caption{The t-SNE of the aggregated embeddings in each block of session "R8001\_M8004\_MS801".}
  \label{fig:tsne}
\end{figure}

\subsection{Results}
Table \ref{tab:results} shows the results of the proposed online TS-VAD system on both the AliMeeting Eval set and Test set, where the block size is 16 seconds, and the block shift is 2 seconds. From the results, we can find that the proposed online TS-VAD system always shows better overall performance than the offline clustering-based system. The reason is that the online TS-VAD model can retrieve the speaker identity from the overlap regions. We also notice that the performance of the online TS-VAD is a little bit worse than the clustering-based method on the 2- and 3-speaker data in AliMeeting Test set. Actually, we find that there is a session that shows very bad performance on each of the 2- and 3-speaker datasets since the number of speakers is not estimated correctly. If these two recordings are excluded for scoring, the online TS-VAD also shows better performance than the clustering-based method on all subsets with different numbers of speakers. Compare with the offline single-channel TS-VAD method of the winner system, our online method does not show much performance degradation.

The DERs with different block size $b$ and block shift $s$ is shown in Figure \ref{fig:results}. When the block size $b$ is fixed, inferring with a smaller block shift $s$ shows better performance most of the time. The reason is that the block with larger block shift may contains more than one new speaker, which does not satisfy our assumption. The model reaches the lowest DER when $b=16$ and $s=2$, which means that latency of our method is 2 seconds, but we need the first 16 seconds for initialization.

Figure \ref{fig:tsne} shows the aggregated embedding in each block during inference. The embeddings from the 4 speakers are well separated even we do not make any assumption about the target speaker embedding during the training stage.

\section{Conclusions}
\label{sec:conclusion}
In this paper, we propose an online TS-VAD method, where the target speaker embeddings are obtained in an online manner. In addition, the TS-VAD model and the speaker embedding extractor share the same front-end, which significantly reduces the computational cost as we do not need to extract the embedding with an additional embedding model. Experimental results show that the proposed method has superior performance to the offline clustering-based system. 

However, the ResNet-based embedding may not be the best choice as the ResNet front-end are not time-efficient for the online task. Besides, we do not make any assumptions about the target speaker embedding during the training stage. In the future, we are going to employ a simpler front-end, e.g., d-vector~\cite{dvector}, which may be more suitable for the frame-level embedding learning in this online scenario. In addition, we will design new objective function on embeddings to obtain a more stable TSEB during the inference stage. Finally, we can extend this method to multi-channel data with some modifications to the model architecture. 

\bibliographystyle{IEEEtran}

\bibliography{mybib}

\end{document}